\documentclass[acmsmall,screen]{acmart}

\usepackage{makecell}
\usepackage{multirow}
\usepackage{colortbl}
\usepackage{subcaption}
\usepackage{wrapfig}

\usepackage{enumitem}
\setlist[itemize]{leftmargin=*}
\setlist[enumerate]{leftmargin=*}

\usepackage[most]{tcolorbox}
\usepackage{xcolor}

\usepackage{graphicx}
\usepackage{subcaption}

\usepackage{listings}
\definecolor{halfgray}{gray}{0.35}
\definecolor{deepblue}{rgb}{0,0,0.5}
\definecolor{deepred}{rgb}{0.6,0,0}
\definecolor{deepgreen}{rgb}{0,0.5,0}
\definecolor{highlightorange}{rgb}{0.98, 0.92, 0.8}

\makeatletter
\newcommand\HL{%
	\gdef\lst@alloverstyle##1{%
		\fboxrule=0pt
		\fboxsep=0pt
		\colorbox{lightgray}{\strut##1}%
	}%
}
\newcommand\HLoff{%
	\xdef\lst@alloverstyle##1{##1}%
}

\definecolor{lightgreen}{rgb}{0.9,1,0.9}
\definecolor{lightred}{rgb}{1,0.9,0.9}

\makeatletter
\newcommand\HLgreen{%
	\gdef\lst@alloverstyle##1{%
		\fboxrule=0pt
		\fboxsep=0pt
		\colorbox{lightgreen}{\strut##1}%
	}%
}
\newcommand\HLgreenoff{%
	\xdef\lst@alloverstyle##1{##1}%
}

\makeatletter
\newcommand\HLred{%
	\gdef\lst@alloverstyle##1{%
		\fboxrule=0pt
		\fboxsep=0pt
		\colorbox{lightred}{\strut##1}%
	}%
}
\newcommand\HLredoff{%
	\xdef\lst@alloverstyle##1{##1}%
}

\definecolor{promptbg}{HTML}{F0F0F0}
\definecolor{responsebg}{HTML}{E8F5E9}

\newtcolorbox{promptbox}{
  colback=promptbg,
  colframe=black!50,
  boxrule=0.5pt,
  arc=4pt,
  left=4pt,
  right=4pt,
  top=4pt,
  bottom=4pt
}

\newtcolorbox{responsebox}{
  colback=responsebg,
  colframe=black!50,
  boxrule=0.5pt,
  arc=4pt,
  left=4pt,
  right=4pt,
  top=4pt,
  bottom=4pt
}

\newcommand{\name}{FlyCatcher}

\newcommand{\validatedCheckers}{334}
\newcommand{\crossValidatedCheckers}{300}
\newcommand{\improvementCrossValidatedCheckers}{2.6x}
\newcommand{\improvementMutants}{5.2x}
\newcommand{\avgTimeToGenerateChecker}{15 seconds}
\newcommand{\avgTokensToGenerateChecker}{183k}

\setcopyright{none}
\renewcommand\footnotetextcopyrightpermission[1]{}
\renewcommand\acmConference[2][0]{}
\begin{document}

\title{FlyCatcher: Neural Inference of Runtime Checkers from Tests}

\author{Beatriz Souza}
\authornote{Work done partly as an intern at Microsoft Research.}
\email{beatrizbzsouza@gmail.com}
\affiliation{%
  \institution{University of Stuttgart}
  \country{Germany}
}

\author{Chang Lou}
\affiliation{%
  \institution{University of Virginia}
  \country{USA}}
\email{lchlou@virginia.edu}

\author{Suman Nath}
\affiliation{%
  \institution{Microsoft Research}
  \country{USA}
}
\email{Suman.Nath@microsoft.com}

\author{Michael Pradel}
\affiliation{%
 \institution{CISPA Helmholtz Center for Information Security}
 \country{Germany}}
\email{michael@binaervarianz.de}
\renewcommand{\shortauthors}{Souza et al.}

\begin{abstract}

  Complex software systems often suffer from silent failures, i.e., violations of the intended semantics that do not cause explicit errors.
  A promising approach to detect such errors is to use system-specific runtime checkers that monitor the execution of a system and check for violations of the intended semantics.
  However, writing such checkers for a given software system is challenging and time-consuming, and hence, rarely done in practice.
  This work presents \name{}, an automated approach to derive runtime checkers from existing tests, i.e., from a resource available for most software systems.
  The critical challenge of such an approach is to generalize the behavioral properties encoded in a test case to arbitrary executions of a system.
  \name{} addresses this challenge through a combination of LLM-based synthesis, static analysis, and dynamic validation, which infers a checker that monitors specific method calls and asserts properties that should hold when they are called.
  The inferred checkers are stateful, i.e., they reason about the system's behavior by maintaining a shadow state that abstracts the actual system state as needed by the checker.
  Our evaluation applies \name{} to 400 tests from four widely used, complex software systems.
  The approach infers \validatedCheckers{} checkers, out of which \crossValidatedCheckers{} are found to be correct via cross-validation.
  Compared with a state-of-the-art approach, our approach infers \improvementCrossValidatedCheckers{} more correct checkers, which enables it to detect \improvementMutants{} more errors.
  By contributing to the automated inference of runtime checkers from tests, this work enables the broader adoption of runtime checking as a practical approach to detect silent failures in complex software systems.

\end{abstract}




\maketitle

\section{Introduction}

Complex software systems suffer from bugs that elude traditional testing and formal verification.
In particular, software systems often experience \emph{silent failures}, i.e., violations of the expected semantics that do not surface as explicit errors.
Despite the lack of overt symptoms, silent failures can corrupt data, produce incorrect results, and introduce security vulnerabilities.
Recent studies report that such failures are prevalent in production deployments, e.g., accounting for 39\% of failures examined by Lou et al.~\cite{lou2022demystifying}.
These failures are notoriously hard to debug as their effects may be detected only much later, long after the root cause has vanished.
Consequently, there is a strong need for techniques that quickly detect silent failures as they occur.

A promising approach to detect such bugs is to deploy \emph{semantic checkers} that run alongside the software and continuously verify that runtime behavior conforms to the intended semantics.
The properties that such checkers should verify are highly domain-specific, e.g., in-order message delivery, correct session timeout handling, or maintaining replication invariants in a distributed system. 
Hence, manually writing semantic checkers typically demands deep domain expertise and painstaking manual effort.
The scale, complexity, and semantic richness of real-world software systems make manual checker construction both challenging and brittle.

Prior work has explored techniques for automatically inferring semantic checkers.
For example, the pioneering Daikon system~\cite{Ernst2001} mines invariants statistically from execution traces.
While useful, such invariants often capture relationships among low-level events rather than high-level semantics that are meaningful to developers.
To address this gap, T2C~\cite{lou2025deriving} proposes deriving semantic checkers from existing tests, i.e., a resource that is anyway available for many software projects.
This kind of approach leverages the insight that the carefully designed workloads and assertions encoded in tests convey rich knowledge of the expected system behavior.
The knowledge captured in tests reflects the developers' intent: tests concretely demonstrate what the system should do and why.
Intuitively, if we can lift these concrete, workload-specific tests into general, runtime checkers, we obtain checkers that are both semantically grounded and operationally useful.

Automatically converting tests into semantic checkers involves multiple challenges, though.
First, tests and their assertions are crafted for specific workloads, whereas runtime checkers must operate on arbitrary executions of the system.
For example, consider a test that creates an ephemeral data structure in a distributed system, assigns a specific name to it, and asserts that the data structure is removed after a timeout.
This test cannot be used verbatim as a semantic checker because other executions of the system may create arbitrarily named data structures.
T2C~\cite{lou2025deriving} uses static analysis to generalize tests into parameterized checkers (e.g., over the data structure name).
However, this generalization is challenging because tests may contain multiple concrete values that must be parameterized and often intermix several assertions whose logic is intertwined with test scaffolding.
The challenge is to disentangle these elements and generalize them correctly.

Second, generating effective checkers requires reasoning about constants and their meaning.
In particular, tests commonly include magic values and domain-specific constants.
Prior work~\cite{lou2025deriving} can fail to account for the role these values play, producing incorrect generalizations.
For example, consider this test:
\texttt{list = new List();\; list.add(1);\; assert(list.length == 1);}

A purely structural analysis is unable to distinguish whether the value \texttt{1} in the assertion is (a) the value added, (b) a constant, or (c) a value tied to internal object state.
As a result, a generated checker might incorrectly tie \texttt{length} to the value added, e.g., asserting \texttt{length == 5} after adding element \texttt{5}, rather than to the count of elements.

Finally, many meaningful properties are stateful.
For example, validating a list's length, the number of active sessions, or the membership of a replica set requires tracking how operations update system state over time.
Designing stateful checkers requires (a) identifying which operations affect state, (b) reasoning about how system operations update state, and (c) maintaining complete coverage of all state-affecting operations.
For example, a property that represents the length of a list must be incremented on \texttt{add()} and decremented on \texttt{delete()} operations.
Without completeness, the state expected by the checker will drift from the true state, producing false positives or missed failures.
Existing approaches do not support such comprehensive stateful checking.

This paper presents \name{}, a novel approach for automatically inferring semantic checkers from existing tests by combining large language model (LLM)-based synthesis, static analysis, and dynamic validation. 
Given a test case as input, the approach generates a semantic checker that captures the high-level intent of the test and can be deployed in production to detect silent failures.
The approach combines lightweight static analysis, e.g., to identify the methods called in the test, with LLM prompting, e.g., to generate candidate checkers.
To increase the correctness of generated checkers, \name{} employs a feedback loop that validates candidates via static analysis and by executing them on a suite of validation tests.
When discrepancies arise, \name{} provides structured feedback to the LLM to refine the checker iteratively.

Unlike prior work, our approach can generate checkers that are substantially more general, robust to diverse workloads, and capable of reasoning about magic values and domain-specific constants.
Our intuition is that LLMs can leverage semantic context -- including code, comments, and identifier names -- to infer why a test asserts what it asserts, rather than merely what it asserts.
The LLM uses this context to synthesize checkers that target high-level semantics rather than low-level event correlations.
This ability enables correct parameterization (e.g., distinguishing element values from the element count of a list) and better generalization beyond a test's original workload.
Another contribution of \name{} is its ability to generate stateful checkers.
Each inferred checker maintains a checker-specific abstraction of the system's state, called the \emph{shadow state}, and uses it to validate stateful properties.
At runtime, the checker continuously compares the observed system state against the shadow state and reports any discrepancies.
Conceptually, a checker that maintains a semantically faithful shadow state functions like a lightweight reference model: Whenever the implementation deviates from this reference, the mismatch signals a semantic failure.


We implement \name{} for Java and evaluate it on 400 test cases from four complex open-source systems.
Given these tests, the approach generates \validatedCheckers{} semantic checkers, showing its ability to support a wide range of testing scenarios.
Cross-validating these checkers against held-out tests, we find \crossValidatedCheckers{} of them to be correct, which increases the number of correctly inferred checkers by \improvementCrossValidatedCheckers{} over the state-of-the-art T2C approach~\cite{lou2025deriving}.
To evaluate the effectiveness of the generated checkers at detecting silent failures, we use mutation testing and finds the \name{}-generated checkers to detect \improvementMutants{} more mutants than T2C-generated checkers.
The costs of using our approach are modest: On average, generating a checker takes \avgTimeToGenerateChecker{} and \avgTokensToGenerateChecker{} LLM tokens (USD 0.60) and running a checker imposes a runtime overhead up to 2.7\%--40.3\% (depending on the target system).

In summary, this paper makes the following contributions:
\begin{itemize}
	\item We present the first LLM-based approach to infer semantic runtime checkers from tests, which results in substantially more general and robust checkers than when using an existing static-analysis-based technique.
	\item We introduce a shadow state mechanism that allows the inferred checkers to maintain a checker-specific abstraction of the system's state and reason about stateful properties.
	\item We implement our ideas in \name{},\footnote{Our implementation is available at \url{https://github.com/biabs1/FlyCatcher}.} evaluate it on four real-world Java systems, and find that it significantly outperforms the best available approach in terms of correctly inferred checkers and bug detection ability.
\end{itemize}

\section{Motivating Example and Problem Definition}
\label{sec:motivation and problem}

The following illustrates the kind of programming error we are addressing with an example based on Zookeeper, a widely used distributed system project, and then defines the problem addressed by our work.
As a motivating example, consider a hypothetical bug in the \texttt{getChildren} method of the \texttt{DataNode} class of Apache Zookeeper, as shown in Figure~\ref{fig:motivation_fault}.
The method is supposed to return a set of children nodes.
However, due to a programming error, it always returns an empty set, regardless of the actual state of the object.
This bug can lead to silent failures where the system behaves incorrectly without crashing or throwing exceptions.

\begin{figure}[t]
  \centering

  \begin{minipage}[c]{0.48\linewidth}
    \centering
    \begin{lstlisting}
public synchronized Set<String> getChildren() {
    if (children == null) {
        return EMPTY_SET;
    }
    (@\HLred@)return Collections.emptySet();(@\HLredoff@)
}
    \end{lstlisting}
    \caption{Example of bug missed in Zookeeper. The highlighted line is the faulty line that causes the method to always return an empty set.}
    \label{fig:motivation_fault}
  \end{minipage}\hfill
  \begin{minipage}[c]{0.48\linewidth}
    \centering
    \begin{lstlisting}
public void testGetChildrenShouldReturnEmptySetWhenThereAreNoChidren() {
    // create DataNode and call getChildren
    DataNode dataNode = new DataNode();
    Set<String> children = dataNode.getChildren();
    assertNotNull(children);
    assertEquals(0, children.size());

    // add child,remove child and then call getChildren
    String child = "child";
    dataNode.addChild(child);
    dataNode.removeChild(child);
    children = dataNode.getChildren();
    assertNotNull(children);
    assertEquals(0, children.size());
}
    \end{lstlisting}
    \caption{Example of test case in Zookeeper.}
    \label{fig:motivation_test}
  \end{minipage}

\end{figure}

As testing is limited to a specific set of inputs and scenarios, such bugs can easily be missed.
Indeed, the bug in Figure~\ref{fig:motivation_fault} remains undetected when running the comprehensive test suite of Zookeeper.
Interestingly, there are existing test cases that check the correctness of the \texttt{getChildren} method, such as the test shown in Figure~\ref{fig:motivation_test}.
The existing test case performs the following steps: 1) Instantiate a \texttt{DataNode} object and call \texttt{getChildren}; 2) Assert that the returned set is non-null and empty; 3) Add and remove a child, then repeat the assertions.
These assertions capture important semantic properties, namely the correctness of \texttt{getChildren} under different states.
However, the test is limited to its specific workload and assertions, and thus cannot catch the bug in Figure~\ref{fig:motivation_fault}.

To generalize the bug detection capability of such tests, we aim to transform them into runtime checkers that monitor and assert the semantic properties encoded in tests, but generalized to work on arbitrary executions of the system.
We call this problem the \emph{test-to-checker} problem, defined as follows:
Given a test $t$ and two sets of additional tests, context tests $T_{context}$ and validation tests $T_{val}$ with $t \notin T_{context} \cup T_{val}$, our goal is to generalize $t$ into a runtime checker $c$.
The runtime checker monitors invocations of state-changing methods exercised by $t$ and validates that the state of the system satisfies the semantic properties encoded in $t$.
The context tests in $T_{context}$ are used by our approach to provide additional semantic information about the system under test, which helps the generalization of $t$ into $c$.
The validation tests in $T_{val}$ are used by our approach to ensure that the generated checker $c$ is not overfitted to $t$ and can generalize to other scenarios.

Addressing this problem involves three key challenges not sufficiently addressed by prior work:

\begin{enumerate}
	\item \emph{Generalizing workloads.} The test $t$ encodes a specific workload, but the checker $c$ should be able to handle arbitrary workloads that may differ from those in $t$.
	By ``workload'' we mean the sequence of method invocations on the system under test and the parameters passed to these methods.
	For the example in Figure~\ref{fig:motivation_test}, the workload hard-codes the number of children added to the \texttt{DataNode} (one) and the name of the child (``child'').
	Instead, we want the checker $c$ to be able to handle arbitrary sequences of adding and removing children with arbitrary names.
	Addressing this challenge requires reasoning about the meaning of magic values and domain-specific constants that appear in the test.

	\item \emph{Generalizing assertions.} The assertions in $t$ are specific to the workload encoded in $t$, but the checker $c$ should be able to validate semantic properties under arbitrary workloads.
	For our example, the checker should validate that the set returned by \texttt{getChildren} is always non-null and that its size matches the number of children added minus the number of children removed, regardless of the specific sequence of additions and removals.
	Clearly, the assertions in $t$ cannot be used verbatim in $c$ as they are tied to the specific workload in $t$.

	\item \emph{Reasoning about internal state.} The semantic properties encoded in $t$ often involve the internal state of the system, which may change over time as methods are invoked.
	To keep track of such state, the checker $c$ needs to maintain its own representation of the relevant internal state and update it as methods are invoked.
	For our example, the checker needs to track the number of children added and removed to correctly validate the size of the set returned by \texttt{getChildren}.
\end{enumerate}
	
Prior work on deriving runtime checking mechanisms does not sufficiently address these three challenges.
One line of prior work, such as Daikon~\cite{Ernst2001} and Dinv~\cite{grant2018inferring}, focuses on mining invariants from program executions.
While these approaches provide a way to derive generalized assertions, they rely on a rich set of execution traces to identify patterns and invariants.
Another line of work infers semantics rules, such as that every invocation of a method $x$ implies a subsequence of invocation of another methods $y$~\cite{lou2022demystifying}.
However that approach requires test cases that reproduce known failures as its input, which limits its applicability.
Approaches like DLint~\cite{issta2015}, DynaPyt~\cite{fse2022-DynaPyt}, and Dylin~\cite{fse2025-Dylin} offer another perspective, by checking for common programming errors at runtime.
However, they focus on generic error patterns and do not check any project-specific properties.
The recent T2C approach~\cite{Lou2025} is most closely related to our work, as it also focuses on generalizing test cases into runtime checkers.
T2C addresses Challenges~1 and~2 via static analysis, which limits its ability to capture the intent of the tested code and the role that magic values and constants play in the test.
Moreover, T2C does not address Challenge~3 because it lacks a mechanism to track and reason about the internal state of the system.
Getting back to our motivating example, given the test in Figure~\ref{fig:motivation_test}, the existing approach fails to catch the bug in Figure~\ref{fig:motivation_fault}.
The reason is that the T2C-generated checker does not track the number of children added and removed, and instead always asserts the size to be zero, which is incorrect.

\section{Approach}

The following presents \name{}, a novel approach to address the \emph{test-to-checker} problem defined in Section~\ref{sec:motivation and problem} through a combination of LLM-based synthesis, lightweight static analysis, and dynamic validation.
We start by providing an overview of the approach (Section~\ref{sec:overview}) and then describe its key components in detail (Sections~\ref{sec:shadow state}--\ref{sec:instrumentation}).

\begin{figure}[t]
  \centering
  \includegraphics[width=0.7\columnwidth]{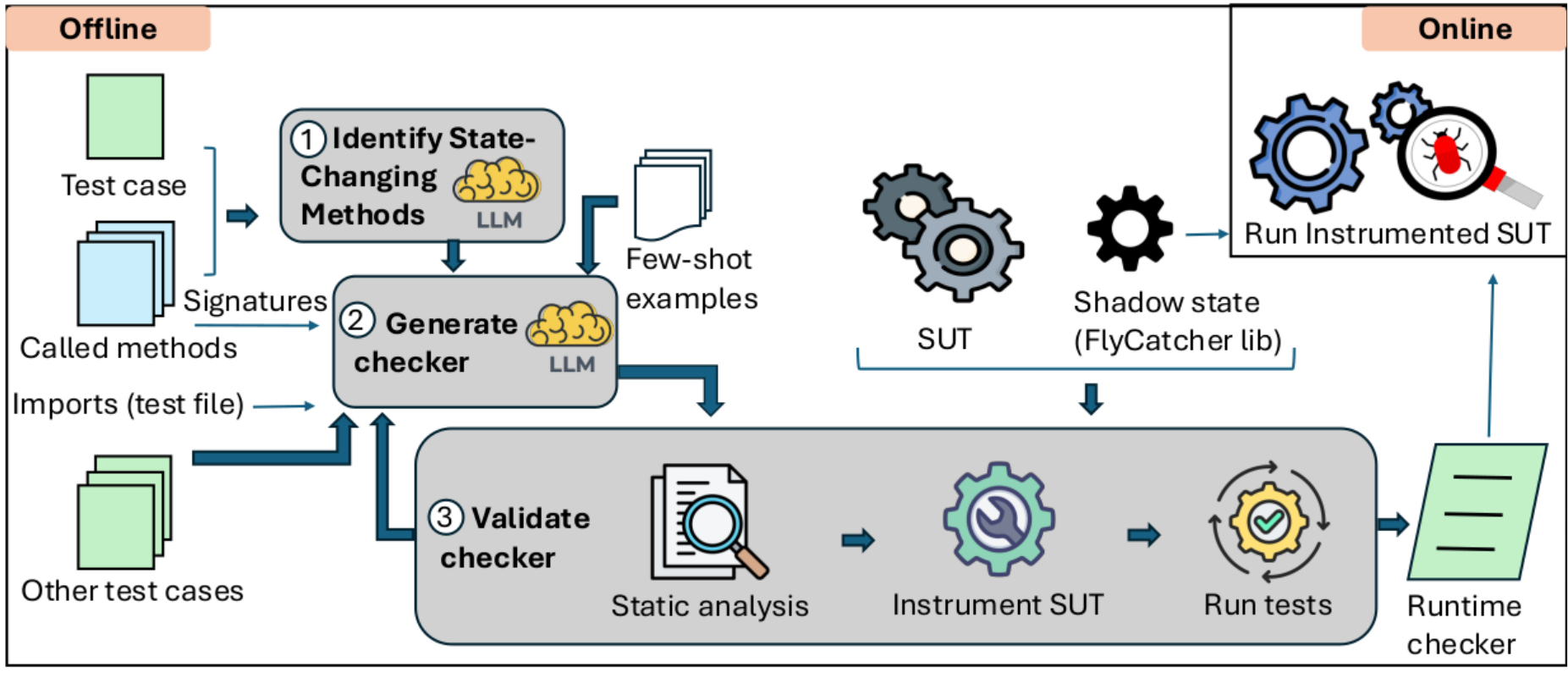}
  \caption{Overview of \name{}.}
  \label{fig:overview}
\end{figure}

\subsection{Overview}
\label{sec:overview}

Figure~\ref{fig:overview} presents an overview of our approach, which consists of two phases: an \textit{offline} phase, during which the approach infers checkers from tests, and an \textit{online} phase, during which the checkers are executed alongside the system.

In the offline phase, \name{} takes a \textit{target test} $t$, a set $T_{context}$ of context tests, and a set $T_{val}$ of validation tests from the system's test suite as input, infers a generalized runtime checker from $t$ and validates it using $T_{val}$.
This phase consists of three main steps, as shown in the gray boxes in Figure~\ref{fig:overview}.
First, \name{} uses an LLM to identify those methods among all methods called in $t$ that affect the state of the system, as well as additional methods that are called in $T_{context}$ and can affect the same system state.
We call these methods the \emph{state-changing methods}.
Second, the approach uses a combination of static analysis and LLM prompting to generate a runtime checker that monitors the state-changing methods and generalizes the properties tested by $t$.
Finally, \name{} validates each runtime checker using a multi-step process, and if necessary, refines the checker by repeating step~2. 
The validation involves lightweight static analysis (e.g., ensuring the checker compiles and contains at least one assertion) and dynamic validation (e.g., running the checker alongside the system and ensuring that it passes all tests in $T_{val}$).
If any validation step reports a problem, an error message is fed back to the checker generation step, and the process repeats until a successful checker is produced or a maximum of \textit{k} attempts is reached.

In the online phase, the validated runtime checker is deployed alongside the system, which \name{} instruments to invoke the checker whenever a state-changing method is called.
When invoked, the checker observes the operation being executed and the current state of the system, updates its own representation of the system state (the \emph{shadow state}, see below), and asserts that the properties encoded in the checker hold.
Runtime checking can be performed on any execution of the system, including executions of additional, held-out tests (as done in our evaluation) or in production.

\subsection{Shadow State}
\label{sec:shadow state}

A key feature of the runtime checkers generated by \name{} is their ability to reason about the internal state of the system under analysis.
To reason about internal system state, the approach maintains a \emph{shadow state}, i.e., an abstraction of the system's relevant state that evolves in parallel with the real system but is isolated from it.
This shadow state enables the checker to reason about and validate expected behaviors without introducing side effects or altering the actual runtime environment.

The shadow state is represented as a hierarchical mapping $S: O \to (P \to V)$, where:
\begin{itemize}
	\item $O$ is the set of objects in the target system,
	\item $P$ is the set of property names, and
	\item $V$ is the set of property values.
\end{itemize}

This representation of the shadow state allows the checker to maintain arbitrary information about the objects in the system, such as the values of their fields or other derived properties.
For example, a checker derived from the test in Figure~\ref{fig:motivation_test} may maintain, for each \texttt{DataNode} object, a property ``children'' that tracks the set of children added and removed via the \texttt{addChild} and \texttt{removeChild} methods.
Another checker may store only the number of children, rather than the children themselves.
Leaving the properties to track open-ended, instead of enforcing a specific structure, allows the inferred checkers to capture a wide range of semantic properties.

To create, maintain, and use the shadow state, checkers perform three main operations: initialization, updates, and reads.
When the checker observes an object $o \in O$ for the first time, it initializes its shadow state by creating a new mapping $S(o)$ that captures the initial values $f_o : P \to V$ of the relevant properties of~$o$.
To maintain the shadow state as the system executes, the checker updates the shadow state whenever it observes a state-changing operation that affects an object~$o$.
Finally, to validate that the properties encoded in the checker hold, the checker reads values from the shadow state and compares them against the actual state of the system.
For example, a checker may assert that the size of the set of children in the shadow state matches the size of the set returned by the \texttt{getChildren} method in the real system.

\subsection{Identifying State-Changing Methods}
\label{sec:identify state-changing}

The idea of the runtime checkers generated by \name{} is to monitor and reason about the state of the system as it evolves over time.
A crucial step toward this end is to identify the methods that can change the state of the system, as these methods are the ones that the checker needs to monitor and handle to keep its shadow state up to date.
Given the target test $t$, \name{} first analyzes the test code to identify the methods that are invoked and may have side effects.
These state-changing methods are then used in the subsequent checker generation (Section~\ref{sec:checker_generation}), validation (Section~\ref{sec:validation}), and instrumentation steps (Section~\ref{sec:instrumentation}).

Identifying state-changing methods is a non-trivial task that can be addressed in various ways.
One possible approach is to statically analyze the code of the methods invoked in the test and to reason about their side effects.
Prior work has proposed analyses to identify pure methods, i.e., methods guaranteed to not have any side effects~\cite{sualcianu2005purity,rountev2004precise}.
Such analyses typically depend on call graph analysis and points-to analysis, which are non-trivial to scale to large code bases.
Another possible approach is to use heuristics based on method names, such as considering methods with names containing words like ``add'', ``remove'', or ``set'' as state-changing~\cite{Lou2025}.
However, this approach may miss methods with less obvious names because any hard-coded set of method names is unlikely to be comprehensive.

\begin{figure}[t]
\centering
\begin{minipage}{\linewidth}
    \begin{promptbox}
    Many methods called in a test may modify the state of the target system, e.g., write and delete.  
    Given a test case and the implementation of the methods called in the test, identify all method calls that can cause side effects in the target system and produce a new version of the test containing the \texttt{// state-changing} comment in each line with method calls that can cause side effects. Constructors are state-changing methods by default.
    
    Method implementations: [implementations]
    
    Test: [test]

  \end{promptbox}
\end{minipage}
\caption{Prompt for identifying state-changing methods.}
\label{fig:test_annotation_prompt}
\end{figure}

Motivated by the limitations of prior approaches, \name{} leverages the capabilities of LLMs to identify state-changing methods based on their names and the context in which they are used.
Specifically, the approach prompts an LLM with a description of the task, the source code of the methods invoked in the test, and the test code itself.
To accurately identify the methods invoked by a test case, we build on the static analysis framework CodeQL~\cite{codeql}.
Figure~\ref{fig:test_annotation_prompt} shows the prompt template, which asks the LLM to annotate state-changing method calls with a specific comment.
For our example test case in Figure~\ref{fig:motivation_test}, the approach properly identifies \texttt{DataNode}, \texttt{addChild}, and \texttt{removeChild} as state-changing operations and annotates the test case accordingly.
Given the response by the LLM, we parse the test code to identify the methods annotated as state-changing.

\subsection{Checker Generation}
\label{sec:checker_generation}

Based on the set of state-changing methods, the next step of \name{} is to generate a runtime checker that monitors invocations of these methods and validates the semantic properties encoded in the target test $t$.
Prior approaches to this problem often rely on specific templates or patterns, such as invariants expressed via binary operations between two variables~\cite{Ernst2001,grant2018inferring} or checks that require a call of method $x$ to be followed by a call of method $y$~\cite{lou2022demystifying}.
To enable \name{} to capture a wider range of semantic properties, we allow checker code to perform arbitrary computations and to include arbitrary assertions.
To synthesize such checkers, \name{} leverages the capabilities of LLMs to ``understand'' the intent of the given target test and generalize it into a runtime checker.
The approach prompts the LLM with a detailed set of guidelines, few-shot examples, the target test $t$, and relevant information about $t$ (e.g., import statements, context tests), as follows.

\subsubsection{Guidelines and Few-Shot Examples}

Because the task of generalizing test cases into runtime checkers is non-standard, it is essential to provide the LLM with detailed guidelines on the expected structure and behavior of the generated checkers.
Our guidelines consist of ten requirements, which we developed and refined after observing runs of \name{} on a small test benchmark:

\begin{enumerate}
  \item The checker is a static and parameterized method.
  \item The checker receives (i) an \texttt{Operation} object, which contains the following attributes: \texttt{signature}, \texttt{baseObject}, \texttt{arguments}, and \texttt{returnValue}; and (ii) a shadow state mapping objects to their properties and respective values.
  \item The checker handles methods that modify the state of object instances. These methods are marked with a \texttt{// state-changing} comment in the target test. For these methods, the checker updates the provided shadow state.
  \item When reading a property from the shadow state, the checker falls back to a default value, as the property may not be in the shadow state yet.
  \item The checker updates the values in the shadow state based on the semantics of the received \texttt{Operation} object.
  \item Toward the end of the checker code, the checker asserts that properties have their expected values. 
  Similar to the assertions in the test case, the assertions may use methods that do not modify the system state. Assert statements should be outside of if-statements, as in the test. Do not insert any return statements.
  \item To obtain the expected value in the assertion, the checker retrieves it from the shadow state.
  \item The checker does not modify the state of the \texttt{baseObject} from the received \texttt{Operation}.
  \item The checker only contains necessary variables and operations, and properly accesses methods and attributes according to their visibility.
  \item The checker code is explained with comments.
\end{enumerate}

To illustrate the requirements, the prompt contains five few-shot examples.
We manually develop these examples, each containing a test case and its corresponding runtime checker.
The few-shot examples are for simple toy classes, such as a \texttt{BankAccount}, a \texttt{ListManager} and a \texttt{Rectangle}, designed to help the LLM understand the task and the expected output format.

\subsubsection{Target Test and Related Information}

The most important input to the LLM is the target test $t$ for which we want to generate a runtime checker.
However, providing the target test alone may be insufficient for the LLM to understand the context and semantics of the code tested in $t$.
For example, this happens when the class under test offers multiple methods but $t$ exercises only a subset of them.
To provide more contextual information on how the system is used beyond the target test, \name{} incorporates existing context tests $T_{context}$ that also test the classes exercised by the target test $t$.
To obtain such context tests, the approach relies on static analysis to identify all tests that invoke methods of the same classes as those invoked in $t$.
As the number of such tests can be large, \name{} randomly samples from them until exceeding a configurable number of tokens.
Another important piece of information for generating a compilable checker is the import statements required by the target test $t$.
The approach hence extracts the import statements from the test file containing $t$ and includes them in the prompt.

Notice that \name{} does not provide the implementation of the tested methods as input to the checker generation. 
This is to avoid that the LLM simply copies the existing implementation into the checker method. 
Instead, \name{} relies on the semantics information conveyed in the tests only, enabling the runtime checkers to be independent of the specific implementation of the tested methods.

\subsubsection{Example of Generated Checker}

\begin{figure}[t]
    \centering
    \begin{lstlisting}
public static void getChildrenChecker(Operation op, Map<Object, Map<String, Object>> shadowState)
{   DataNode baseObject = (DataNode) op.baseObject;
    Map<String, Object> objectState = shadowState.getOrDefault(baseObject, new HashMap<>());

    // Initialize children set if not present
    if (!objectState.containsKey("children")) {
        // Update the objectState from the shadow shadowState
        objectState.put("children", new HashSet<String>());
    }
    // Handle "addChild" operation, which modifies the state of baseObject
    if (op.signature.equals("org.apache.zookeeper.server.DataNode.addChild(java.lang.String)")) {
        // Update the objectState from the shadow shadowState
        Set<String> children = (Set<String>) objectState.get("children");
        children.add((String) op.arguments[0]);
        objectState.put("children", children);
    }
    // Handle "removeChild" operation, which modifies the state of baseObject
    else if (op.signature.equals("org.apache.zookeeper.server.DataNode.removeChild(java.lang.String)")) {
        // Update the objectState from the shadow shadowState
        Set<String> children = (Set<String>) objectState.get("children");
        children.remove((String) op.arguments[0]);
        objectState.put("children", children);
    }

    // Assert that the baseObject returns the correct children set
    // getChildren does not modify the state of baseObject
    Set<String> actualChildren = baseObject.getChildren();
    Set<String> expectedChildren = (Set<String>) objectState.get("children");
    
    // Check that children set is not null
    assertNotNull(actualChildren);
    // Check that the size matches the expected size
    assertEquals(expectedChildren.size(), actualChildren.size());
    // Check that all expected children are present
    for (String child : expectedChildren) { assertTrue(actualChildren.contains(child)); }
  
    // Update the shadow state
    shadowState.put(baseObject, objectState);
}
    \end{lstlisting}
    \caption{Checker generated by \name{} for the motivating example.}
    \label{fig:checker4motivatingexample}
\end{figure}

Figure~\ref{fig:checker4motivatingexample} shows a runtime checker generated by \name{} for the target test in Figure~\ref{fig:motivation_test}.
The checker will eventually be invoked whenever one of the state-changing methods \texttt{addChild} or \texttt{removeChild} is called on a \texttt{DataNode} object.
The checker receives as argument an operation, which represents the current method being invoked, and the shadow state.
First, the checker gets the current state, both from the system's actual state and the shadow state, of the base object of the received operation.
Then, the checker initializes the \texttt{children} value in the shadow state in case it does not yet exist.
After that, the checker modifies the shadow state depending on the operation that was received as the argument.
If the received operation is \texttt{addChild}, the checker adds the child to the shadow state.
Similarly, if the received operation is \texttt{removeChild}, the checker removes the child from the shadow state.
In the end, regardless of the received operation, the checker asserts that the values in the real execution are as expected by comparing them to the values in the shadow state.

\subsection{Checker Validation}
\label{sec:validation}

\begin{figure}[t]
\centering
\begin{minipage}{\linewidth}
    \begin{promptbox}
    When trying to [compile|instrument|execute] the provided checker, the following error happens:
    
    [error]
    
    Please, provide a fixed version of the provided checker to fix the error.
  \end{promptbox}
\end{minipage}
\caption{Checker refinement prompt.}
\label{fig:feedback_prompt}
\end{figure}

\begin{table}[t]
    \footnotesize
    \caption{Validation and feedback for checker refinement.}
    \label{tab:refinement}
    \begin{tabular}{@{}p{13em}p{25em}@{}}
      \toprule
      Condition & Feedback                                                                                                           \\
      \midrule
      \emph{Static validation:} \\
      \midrule
      Syntax error &  Syntax error in Java code. Make sure that the checker method is indeed a single method, i.e. do not output helper methods or classes.                  \\
      No assertion       & The checker does not contain a call to an assertion method. Make sure to include assertions outside comments. \\
      Non-SUT method        & The system under test (SUT) does not contain the following methods: [methods signatures]. Make sure that the checker handles methods from the system under analysis rather than built-in functions or methods from the test suite. \\         
      Non-fully qualified signature & The checker handles methods without fully qualified signature: [unqualified signature]. Use fully qualified names for the method and all argument types. \\
      \midrule
      \emph{Dynamic validation:} \\
      \midrule
      Test failure & The following tests fail: [tests logs]. The checker should be generic and robust enough to meaningfully satisfy all test cases.             \\                                                                 
      Calls state-changing method & This checker is calling a state-changing method. This is not allowed. \\
      \textgreater 30m execution & The checker is making the tests run for more than 30min. \\
      \bottomrule
    \end{tabular}
\end{table}

LLMs, while powerful, can produce incorrect or suboptimal outputs.
To improve the quality of the generated checkers, \name{} uses a multi-step process to validate checkers before returning them to users. 
If and only if a checker passes all the validation steps, \name{} outputs the checker.
Otherwise, the approach uses the prompt in Figure~\ref{fig:feedback_prompt} with one of the feedback messages in Table~\ref{tab:refinement} to produce a fixed version of the checker.
This process is repeated until a successful checker is generated or a number \textit{k} of attempts is reached.
The refinement follows a conversation style, where the LLM has access to the previous versions of the checker and the feedback messages, enabling the model to learn from its own mistakes.
The following explains the validation steps, grouped into static and dynamic validation.

\subsubsection{Static Validation of Checkers}
\label{sec:static_validation}

The first group of validation steps uses lightweight static analysis to check the syntactic correctness and basic properties of the generated checker (see first part of Table~\ref{tab:refinement}).
First, the approach ensures that the generated checker is syntactically correct and can be parsed.
If the LLM outputs a checker method that contains syntax errors or that contains multiple methods instead of a single method, the approach rejects the checker and feeds the error message back to the LLM for refinement. 
Second, \name{} validates that the checker contains at least one call to an assertion method.
The rationale is that a checker without assertions would not be useful, as it does not check any property.
Third, the approach ensures that the checker handles only methods that belong to the system under analysis.
This step is motivated by the observation that LLMs sometimes hallucinate methods that do not exist in the system or try to monitor built-in methods from the Java standard library.
Finally, \name{} checks that the checker code identifies methods using their fully qualified signature. 

\begin{figure}[t]
  \centering

  \begin{minipage}[c]{0.48\linewidth}
    \centering
    \begin{lstlisting}
(@\HLgreen@)package flycatcher.checkers;(@\HLgreenoff@)

(@\HLgreen@)import flycatcher.util.Operation;(@\HLgreenoff@)
(@\HLgreen@)import flycatcher.util.ShadowState;(@\HLgreenoff@)
import org.apache.zookeeper.server.DataNode;
import java.util.*;
import static org.junit.Assert.*;

(@\HLgreen@)public class Checker326 {(@\HLgreenoff@)
    public static void getChildrenChecker(Operation op, Map<Object, Map<String, Object>> shadowState) {
        (@\HLgreen@)if (flycatcher.util.ShadowState.inChecker) (@\HLgreenoff@)
           (@\HLgreen@)throw new java.lang.RuntimeException("Checker is calling a state-changing method.");(@\HLgreenoff@)
        (@\HLgreen@)flycatcher.util.ShadowState.inChecker = true;(@\HLgreenoff@)
        (@\HLgreen@)try {(@\HLgreenoff@)
            // <body of checker method generated by LLM>
        (@\HLgreen@)} finally {(@\HLgreenoff@)
         (@\HLgreen@)   flycatcher.util.ShadowState.inChecker = false;(@\HLgreenoff@)
        (@\HLgreen@)}(@\HLgreenoff@)
    }
(@\HLgreen@)}(@\HLgreenoff@)
    \end{lstlisting}
    \caption{Scaffolding (highlighted in green) added around the LLM-generated checker method.}
    \label{fig:post_processing_example}
  \end{minipage}\hfill
  \begin{minipage}[c]{0.48\linewidth}
        \begin{lstlisting}
public synchronized boolean addChild(String child) {
    (@\HLgreen@)Boolean returnValue = null; (@\HLgreenoff@)
    (@\HLgreen@)try { (@\HLgreenoff@)
        // <body of the original addChild method> 
        // returnValue receives the value that would be returned by addChild
    (@\HLgreen@)} finally { (@\HLgreenoff@)
         (@\HLgreen@)Checker326.getChildrenChecker(new Operation("org.apache.zookeeper.server.DataNode.addChild(java.lang.String)",
             (@\HLgreen@)this, new Object[]{child}, returnValue),ShadowState.state);(@\HLgreenoff@)
    (@\HLgreen@)} (@\HLgreenoff@)
    return returnValue;
}
    \end{lstlisting}
    \caption{Example of instrumentation performed by \name{}.}
    \label{fig:instrumentation_example}
  \end{minipage}
\end{figure}

To turn the checker method generated by the LLM into a compilable and executable checker, \name{} post-processes the generated code by integrating it into a scaffolding file.
Figure~\ref{fig:post_processing_example} illustrates the scaffolding file for the checker in Figure~\ref{fig:checker4motivatingexample}.
The code highlighted in green is part of the scaffolding file, while the rest is the checker code generated by the LLM.
The scaffolding code maintains a flag to ensure that the checker does not call state-changing methods, which would lead to infinite recursion when the instrumented state-changing method calls the checker and the checker calls that method again.
To identify and avoid this problem, \name{} uses a flag called \texttt{inChecker}, which is set to \texttt{true} while the checker code is executing and set to \texttt{false} otherwise.


\subsubsection{Dynamic Validation of Checkers}

Once a checker passes the static validation steps, \name{} performs dynamic validation to ensure that the checker behaves correctly when executed alongside the system (see second part of Table~\ref{tab:refinement}).
The dynamic validation consists of two steps:
(i) instrumenting the system under analysis to invoke the checker when one of the state-changing methods is called (explained below in Section~\ref{sec:instrumentation}) and
(ii) running the provided validation tests $T_{val}$ on the instrumented system.

If the checker is correct and general enough, it should not cause any of the validation tests to fail.
If any of the validation tests fails, \name{} rejects the checker and feeds the error message back to the LLM for further refining the checker.
Beyond test failures, \name{} also checks two additional conditions during the execution of the validation tests.
The first condition is to ensure that the checker does not call any state-changing methods, which would lead to infinite recursion.
By using the scaffolding structure illustrated in Figure~\ref{fig:post_processing_example}, \name{} can detect when a checker calls a state-changing method, and hence, would recurses into itself.
The second condition is to ensure that the execution of the validation tests with the checker activated does not take an excessive amount of time.
Specifically, \name{} considers a checker invalid if the execution of the validation tests with the checker activated takes more than 30 minutes to run.

If the checker passes all dynamic validation steps, \name{} outputs the checker.
For our running example, the checker in Figure~\ref{fig:checker4motivatingexample} passes all static and dynamic validation steps.

\subsection{Instrumentation and Online Checking}
\label{sec:instrumentation}

To be able to monitor the system, \name{} executes the inferred runtime checkers while the system under analysis is running.
The key component to enable this online checking is to instrument the system under analysis to invoke the checker whenever one of the state-changing methods is called.
The instrumentation is performed as a source-to-source transformation that wraps the body of each state-changing method into code that invokes the corresponding checker.
In addition, \name{} also adds to the instrumented source files the necessary import statements to use the checker and the \texttt{Operation} class, which encapsulates information about the current operation being executed, and the \texttt{ShadowState} class.
The instrumentation preserves method signatures, allowing drop-in replacement of the original code.

Figure~\ref{fig:instrumentation_example} shows the instrumented version of the \texttt{addChild} method called in the test in Figure~\ref{fig:motivation_test} and handled by the checker in Figure~\ref{fig:checker4motivatingexample}.
The code highlighted in green is the instrumentation added by \name{}.
To ensure that the checker is called even if the instrumented method raises an exception, \name{} wraps the method body in a \texttt{try-finally} block.
After the state-changing method returns, the instrumentation calls the checker, passing it an \texttt{Operation} object that encapsulates information about the current operation, as well as the shadow state.
The \text{Operation} object contains the fully qualified signature of the state-changing method, the base object of the call, i.e \texttt{this}, and the arguments and return value of the \texttt{addChild} method.

Given the instrumented system, \name{} can perform online checking by executing the system as usual.
Executions can be either driven by additional tests or by real users in production.
Whenever a checker observes a violation of the properties encoded in it, it raises an assertion error, which can cause the system to crash, be logged for later analysis or handled in real-time, depending on the needs of the user.

Getting back to our running example, recall the bug in Figure~\ref{fig:motivation_fault}, which causes the \texttt{getChildren} method to return an empty set even after a child has been added.
When executing the buggy version of Zookeeper along with the inferred checker in Figure~\ref{fig:checker4motivatingexample}, the checker identifies the bug after the first call to \texttt{addChild}. In this case, the \texttt{expectedChildren.size()} in the shadow state is one, but the \texttt{actualChildren.size()} in the original system is zero.

\section{Evaluation}


Our evaluation addresses the following research questions:
\begin{itemize}
  \item RQ1: How effective is \name{} at inferring runtime checkers?
  \item RQ2: Are the checkers generated by \name{} useful to detect errors?
  \item RQ3: Do \name{}'s checkers incur false positives?
  \item RQ4: How does feedback refinement contribute to \name{}'s effectiveness?
  \item RQ5: What are the costs of creating and using runtime checkers?
  
\end{itemize}

\subsection{Experimental Setup}

\paragraph{Target Systems and Tests}

\begin{table}[t]
  \footnotesize
  \caption{Target systems and tests.}
  \label{tab:tests}
  \begin{tabular}{llrlrlrlrrr}
  \toprule
  System               & Version              & \multicolumn{9}{c}{Tests}                                                                                                                                                               \\ \cmidrule{3-11} 
  \multicolumn{1}{l}{} & \multicolumn{1}{l}{} & \multicolumn{1}{c}{All} &  & \multicolumn{1}{c}{W/ SUT calls} &  & \multicolumn{1}{c}{W/ assert} &  & \multicolumn{1}{c}{Passing} & \multicolumn{1}{c}{} & \multicolumn{1}{c}{Targeted} \\
  \midrule
  Zookeeper            & 3.4.11               & 439                     &  & 406                              &  & 295                           &  & 249                         &                      & 100                          \\
  Cassandra            & 3.11.5               & 3,294                    &  & 2,293                             &  &  1,162                         &  & 255                            &                   & 100                             \\
  HDFS                 & 3.2.2                & 4,320                    &  & 2,916                             &  & 2,162                          &  & 1,325                        &                      & 100                            \\
  HBase                & 2.4.0                & 4,287                    &  & 2,869                            &  & 2,257                          &  & 1,812                        &                      &  100                            \\ 
  \midrule
  Sum & & 12,340 & & 8,484 & & 5,876 & & 3,641 & & 400 \\
  \bottomrule
  \end{tabular}
\end{table}

We apply our approach to four complex and widely-used systems, listed in Table~\ref{tab:tests}.
We select these systems because they are representative of real-world software, with complex functionalities and rich APIs, because they contain extensive tests suites, and because they were used to evaluate the most closely related prior work~\cite{Lou2025}.
To select target tests to infer runtime checkers from, we start from all the tests in each system and apply three filtering steps that keep tests with the following properties:
(i) the test calls at least one method from the target system (system under test, SUT);
(ii) the test contains at least one assertion; and
(iii) the test terminates and passes within a three-minute timeout.
This filtering results in 3,641 tests across the four systems, from which we randomly sample 100 from each system to obtain our benchmark of 400 target tests.

\textit{Baseline}.
We compare \name{} with T2C~\cite{Lou2025}, the state-of-the-art approach to derive runtime checkers from tests.
Unlike \name{}, which combines LLM-based reasoning with static and dynamic analysis, T2C is based on static and dynamic analysis only.
We apply their approach to the same 400 target tests as \name{}.

\textit{LLMs}.
As the core of \name{} relies on LLMs, we evaluate our approach with two different models to assess its robustness to the choice of model: Claude Sonnet~4\footnote{https://docs.claude.com/en/docs/about-claude/models/overview}, developed by Anthropic, and GPT-4o\footnote{https://platform.openai.com/docs/models/gpt-4o}, developed by OpenAI.

\textit{Parameters}.
We set the maximum number \textit{k} of attempts to fix a checker to 125.
However, when the same kind of problem, e.g., a compilation error, still persists after five fix attempts in a row, we configure \name{} to stop.
For each query to an LLM, we set the number of completions to one.
Moreover, we limit the set $T_{context}$ of context tests to 30k tokens.

\textit{Implementation and Hardware}.
We implement \name{} mostly in Python. 
To identify the static data used as input for the two first steps of \name{}, we rely on CodeQL~\cite{codeql}.
The system instrumentation is implemented in Java using the Javassist bytecode editing library.
The checker generation and validation experiments are done in a server with 4-core 2.60GHz CPUs, with 15 GB memory, running Ubuntu 24.04.
The remaining experiments, e.g., execution of tests to get the 400 target tests benchmark, mutation analysis, and overhead measurements, are done in a server with 48-core 2.20GHz CPUs, with 251 GB memory, running Ubuntu 22.04.

\subsection{RQ1: Effectiveness}

To assess the effectiveness of \name{} at generating runtime checkers, we use two metrics:
(i) the number of \emph{validated} checkers, i.e., checkers that pass both static and dynamic validation; and
(ii) the number of \emph{cross-validated} checkers, i.e., validated checkers that also pass dynamic validation when executed under workloads different from those used to generate them.
For cross-validation, we run the entire test suite of the target system with all validated checkers enabled.

Table~\ref{tab:checkers} presents the results for \name{}, both with Claude Sonnet~4 and GPT-4o, and for the T2C baseline.
\name{}, when used with either model, produces a substantial number of validated checkers for the target tests.
The Claude Sonnet~4 model is more effective than GPT-4o, producing a total of \validatedCheckers{} validated checkers across all 400 target tests.
We do not report the number of validated checkers produced by T2C at this stage because their approach reports checkers on a per-assertion basis, whereas \name{} generates one checker per test, making the numbers incomparable.
The last column of Table~\ref{tab:checkers} reports the number of cross-validated checkers, which is the most important metric, as it measures to what extent the checkers correctly generalize to new workloads.
We find that \name{} consistently and substantially outperforms the baseline, irrespective of the model used.
While T2C produces only 13--30 cross-validated checkers per system, \name{} with Claude Sonnet~4 yields 69--80 such checkers.
In total, \name{} produces \crossValidatedCheckers{} cross-validated checkers, which is \improvementCrossValidatedCheckers{} more than T2C.

\begin{table}[t]
  \footnotesize
  \caption{Effectiveness of \name{} and baseline at generating checkers.}
  \label{tab:checkers}
  \setlength{\tabcolsep}{1.5pt}
  \begin{tabular}{cclrrr}
  \toprule
  System                     & Target tests         & \multicolumn{1}{c}{Approach} & \multicolumn{3}{c}{Checkers}                                                               \\ \cmidrule{4-6} 
  \multicolumn{1}{l}{}       & \multicolumn{1}{l}{} &                              & \multicolumn{1}{c}{Validated} & \multicolumn{1}{l}{} & \multicolumn{1}{c}{Cross-Validated} \\
  \midrule
  \multirow{3}{*}{Zookeeper} & \multirow{3}{*}{100} & FlyCatcher (Claude Sonnet~4) & \textbf{87}                   &                      & \textbf{80}                         \\
                            &                      & FlyCatcher (GPT-4o)           & 67                            &                      & 61                                  \\
                            &                      & T2C                          & -                            &                      & 30                                  \\ \midrule
  \multirow{3}{*}{Cassandra} & \multirow{3}{*}{100} & FlyCatcher (Claude Sonnet~4) & \textbf{78}                     &                      & \textbf{77}                           \\
                            &                      & FlyCatcher (GPT-4o)           &  63                             &                      &  59                                   \\
                            &                      & T2C                          &  -                             &                      &  13                                   \\ \midrule
  \multirow{3}{*}{HDFS}      & \multirow{3}{*}{100} & FlyCatcher (Claude Sonnet~4) & \textbf{79}                   &                      & \textbf{69}                           \\
                            &                      & FlyCatcher (GPT-4o)           & 61                              &                      & 53                                    \\
                            &                      & T2C                          & -                              &                      &  18                                   \\ \midrule
  \multirow{3}{*}{HBase}     & \multirow{3}{*}{100} & FlyCatcher (Claude Sonnet~4) & \textbf{90}                  &                      & \textbf{74}                           \\
                            &                      & FlyCatcher (GPT-4o)           & 65                            &                      &  58                                   \\
                            &                      & T2C                          & -                            &                      &  27                                   \\ \bottomrule
  \end{tabular}
\end{table}

\subsection{RQ2: Usefulness for Detecting Errors}

As most techniques to find bugs can find only a small subset of all bugs~\cite{ase2018-study}, a meaningful evaluation of \name{}'s ability to find bugs requires a large set of known bugs.
However, real-world bugs are relatively rare and often hard to reproduce.
We thus follow the established practice~\cite{Bacchelli2008,Serra2019,Souza2020} of using mutations~\cite{DeMillo1978mutation,Jia2011} to create a diverse set of known bugs and then measure how many of them our approach detects.
The idea behind mutation testing is to apply small syntactic changes to the original system to create faulty programs called mutants.
A mutant is said to be killed if a test case fails when executed against it.
There is a strong correlation between the ability to detect mutants and real errors~\cite{Just2014mutants}.

The computational cost of using mutation testing is high, due to the high number of generated mutants and the high computing time to execute the test suite against each mutant.
To balance the costs of mutation testing, we focus on Zookeeper in this research question, as it takes ``only'' 48 hours to run on a machine with 40 CPU cores, which is less time than the other systems.
To create mutants, we use PITest~\cite{Coles2016}, a mutation testing tool for Java.
We use all the mutation operators available in PITest, and apply them to all classes containing methods called in those target tests of Zookeeper where \name{} infers a cross-validated checker.
For each mutant that is covered by the target tests but not killed by them, i.e., a bug that would usually remain undetected, we apply the \name{}-generated checkers and measure whether the mutant gets killed when running the target tests with the checkers enabled.
To compare with the baseline, we also run T2C's checkers on the same set of mutants.
However, as their approach is not fully automated and requires to manually configure the classes to be instrumented for each run, we apply T2C only to those mutants that \name{} is able to kill.

Table~\ref{tab:mutants} presents the number of mutants covered, killed, and survived by the target tests.
Out of all 3,366 mutants, 1,822 survive the target tests despite being covered by them.
That is, the assertions in the target tests are not strong enough to catch these mutants.
Running the same target tests, but now with the 80 \name{}-generated and cross-validated checkers enabled, kills 26 of these mutants.
This shows that the checkers produced by \name{} are indeed useful to detect errors that would otherwise remain undetected.
Applying checkers generated by the T2C baseline to these 26 mutants, we find that those checkers kill only 5 of them.
That is, \name{} detects \improvementMutants{} more mutants than T2C, which is a substantial improvement.
To further put these numbers in perspective, we refer to a study showing that static bug detectors, i.e., a class of bug detection tools that is widely used in practice, typically find between 1\%--3\% of all bugs in a system~\cite{ase2018-study}.

\begin{figure}[t]
    \centering
    \begin{minipage}[t]{0.25\textwidth}
      \footnotesize
      \centering
    \captionof{table}{Usefulness for finding errors.}
    \label{tab:mutants}
    \begin{tabular}{lr}
      \toprule
      \multicolumn{2}{c}{Mutants covered by target tests:} \\ \midrule
      All                            & 3,366               \\
      Killed by target tests         & 1,544               \\
      Survived                       & 1,822               \\ \midrule
      \multicolumn{2}{l}{Killed by checkers:}              \\ \midrule
      FlyCatcher                     & \textbf{26}        \\
      T2C                            & 5                  \\
      \bottomrule
    \end{tabular}
    \end{minipage} 
    \hfill
    \begin{minipage}[t]{0.3\textwidth}
      \footnotesize
      \centering
    \captionof{table}{Reasons why \name{} misses some bugs.}
    \label{tab:undetection}
    \begin{tabular}{lr}
    \toprule
    Reason & Nb.\ cases \\ \midrule
    Limited workload & 12 \\
    Equivalent mutant & 18 \\ \midrule
    Total & 30 \\
    \bottomrule
    \end{tabular} 
    \end{minipage}
    \hfill
    \begin{minipage}[t]{0.35\textwidth}
        \centering
        \begin{lstlisting}
public long getOpenFileDescriptorCount() {
  Long ofdc;
  if (!ibmvendor) {
    ofdc = getOSUnixMXBeanMethod("getOpenFileDescriptorCount");
    (@\HLred@)return (false ? ofdc.longValue() : -1);(@\HLredoff@)
  }
  // ...
}
      \end{lstlisting}
        \captionof{figure}{Bug detected by \name{}: The equality check is replaced with \texttt{false}, causing the method to return \texttt{-1} when \texttt{bmvendor} is false.}
        \label{fig:mutant}
      \end{minipage}
\end{figure}

\textit{Examples of detected and undetected mutants}.
The faulty version of the {\tt getChildren} method in Figure~\ref{fig:motivation_fault} is one of the mutants generated by PITest with the \textit{Empty returns} mutator, which is detected by \name{}.
Figure~\ref{fig:mutant} shows another mutant detected by \name{}, where the buggy code never returns the result of evaluating \texttt{ofdv.longValue()}, as the condition in the ternary expression is \texttt{false}.
\name{} produces a checker that monitors the values returned by \texttt{getOpenFileDescriptorCount()} and is able to detect this bug.

To better understand the limitations of our approach, we manually inspect 30 of the mutants that are not killed by \name{}. 
Table~\ref{tab:undetection} shows the two reasons we observe for why \name{} misses these bugs.
12 of the inspected cases are due to \textit{Limited workload}, which means that the existing tests do not trigger the buggy path.
We rely on the existing tests in Zookeeper to run our checkers on the mutants, which may not cover all possible scenarios.
Additionally, the environment we use for our experiments also may not cover all possible scenarios.
The mutant in Figure~\ref{fig:mutantm1} is not detected as our experiments are executed on Linux, and the condition to check for Windows is replaced with false.
The second reason we found for \name{} missing some bugs is \textit{Equivalent mutants}, which accounts for the other 18 cases.
An equivalent mutant is a mutant that is syntactically different from the original program but semantically equivalent. 
Figure~\ref{fig:mutantm2} shows an example of an equivalent mutant missed by \name{}.
In this case, the initial assignment to the long variable is removed. However, in Java, long variables receive zero as default value.
Figure~\ref{fig:mutantm3} shows another case we found in this category.
The mutant modifies the argument passed when initializing a \texttt{HashSet}.
However, this does not change the functional correctness of the program, as the argument determines only the initial capacity of the set, which can grow dynamically.

\begin{figure}[t]
  \centering

  \begin{subfigure}[b]{0.32\linewidth}
    \centering
    \begin{lstlisting}
public boolean getUnix() {
  (@\HLred@)if (false) {(@\HLredoff@)
    return false;
  }
  return (ibmvendor ? linux : true);
}
    \end{lstlisting}
    \caption{The condition \texttt{Windows} is replaced with \texttt{false}.}
    \label{fig:mutantm1}
  \end{subfigure}\hfill
  \begin{subfigure}[b]{0.32\linewidth}
    \centering
    \begin{lstlisting}
/** 
* these are the number of acls that we have in the datatree 
*/
(@\HLred@)- long aclIndex = 0;(@\HLredoff@)
(@\HLgreen@)+ long aclIndex;(@\HLgreenoff@)
    \end{lstlisting}
    \caption{Equivalent mutant where the assignment to \texttt{aclIndex} is removed.}
    \label{fig:mutantm2}
  \end{subfigure}\hfill
  \begin{subfigure}[b]{0.32\linewidth}
    \centering
    \begin{lstlisting}
// let's be conservative on the typical number of children
(@\HLred@)children = new HashSet<String>(9);(@\HLredoff@)
    \end{lstlisting}
    \caption{Equivalent mutant where the HashSet size argument changes from \texttt{8} to \texttt{9}.}
    \label{fig:mutantm3}
  \end{subfigure}

  \caption{Examples of mutants missed by \name{}.}
  \label{fig:undetectedMutants}
\end{figure}

\subsection{RQ3: False Positives}

To measure the false positives of \name{}'s checkers in real executions, we follow the steps to calculate false positives done in T2C:
run the system under analysis on five nodes using Jepsen~\cite{jepsen}, a widely used testing framework for distributed systems.
Jepsen generates common workloads and automatically injects network faults every 10 seconds.
As setting up this experiment requires non-trivial configuration efforts, we focus on the two systems for which \name{} generates the highest number of cross-validated checkers: Zookeeper and Cassandra.
For each checker, we run the system with Jepsen for five minutes. 
Table~\ref{tab:falsePositives} shows the results. In Zookeeper, 44\% of the 80 checkers are activated and, in total, called over five million times, from which only 169 calls failed.
In Cassandra, 37\% of the 77 checkers are activated and, in total, called 37,600 times, all of which succeed.
We conclude that \name{} produces checkers with very few false positives, making the checkers suitable for use in practice.

\begin{table}[]
  \footnotesize
  \caption{Analysis of false positives caused by \name{}'s checkers.}
  \label{tab:falsePositives}
  \begin{tabular}{crrrr}
    \hline
    System    & \multicolumn{1}{c}{Activated checkers} & \multicolumn{1}{c}{Checker calls} & \multicolumn{1}{c}{Checker failures} & \multicolumn{1}{c}{False positive} \\ \hline
    Zookeeper & 44\%                                   & 5,509,724                           & 169                                  & 0.00003                            \\
    Cassandra & 37\%                                   & 37,600                             & 0                                    & 0                                  \\ \hline
  \end{tabular}
\end{table}

\subsection{RQ4: Ablation Study on Feedback-Based Refinement}

A core component of \name{} is the feedback-based refinement of checkers, which leverages both static and dynamic validation to iteratively fix invalid checkers (Section~\ref{sec:validation}).
We evaluate the contribution of each validation step to \name{}'s overall effectiveness.
To this end, we measure the number of validated checkers generated by \name{} under different feedback configurations for all 400 target tests drawn from the four systems, using both models.

The results are summarized in Table~\ref{tab:feedback}.
Without any feedback, \name{} produces only 48 validated checkers with Claude Sonnet~4 and 18 with GPT-4o. Incorporating static validation feedback significantly improves effectiveness, yielding 190 and 131 validated checkers with Claude Sonnet~4 and GPT-4o, respectively. Finally, dynamic validation feedback further improves the effectiveness, resulting in 334 and 256 validated checkers for Claude Sonnet~4 and GPT-4o, respectively. These findings demonstrate that the feedback mechanism in \name{} is critical to its effectiveness, regardless of the underlying model.

\begin{table}[t]
  \footnotesize
  \caption{Effectiveness of \name{} with different kinds of feedback provided to refine checkers.}
  \label{tab:feedback}
\begin{tabular}{cclrrr}
\toprule
System                     & Target tests         & Approach                          & \multicolumn{3}{c}{Validated checkers}                                                  \\ \cmidrule{4-6} 
\multicolumn{1}{l}{}       & \multicolumn{1}{l}{} &                                   & \multicolumn{1}{l}{Claude Sonnet~4} & \multicolumn{1}{l}{} & \multicolumn{1}{l}{GPT-4o} \\
\midrule 
\multirow{3}{*}{Zookeeper} & \multirow{3}{*}{100} & FlyCatcher w/o feedback           &   8                                  &                      &  9                          \\
                           &                      & FlyCatcher w/ static validation &   49                                  &                      & 37                           \\
                           &                      & FlyCatcher w/ dynamic validation  &   \textbf{87}                                  &             &  \textbf{67}                          \\ 
\midrule
\multirow{3}{*}{Cassandra} & \multirow{3}{*}{100} & FlyCatcher w/o feedback           &   12                                  &                      &  4                          \\
                           &                      & FlyCatcher w/ static validation &   47                                  &                      &  26                          \\
                           &                      & FlyCatcher w/ dynamic validation  &   \textbf{78}                                  &         &    \textbf{63}                        \\
\midrule
\multirow{3}{*}{HDFS}      & \multirow{3}{*}{100} & FlyCatcher w/o feedback           &   12                                  &                      &  2                          \\
                           &                      & FlyCatcher w/ static validation &   48                                  &                      &  33                          \\
                           &                      & FlyCatcher w/ dynamic validation  &   \textbf{79}                                  &            &  \textbf{61}                          \\ 
\midrule
\multirow{3}{*}{HBase}     & \multirow{3}{*}{100} & FlyCatcher w/o feedback           &  16                                   &                      & 3                           \\
                           &                      & FlyCatcher w/ static validation &  46                                   &                      &  35                          \\
                           &                      & FlyCatcher w/ dynamic validation  &  \textbf{90}                                   &          &   \textbf{65}                         \\
\bottomrule
\end{tabular}
\end{table}

\subsection{RQ5: Costs of the Approach}

\begin{figure}[t]
    \centering
    \begin{subfigure}[b]{0.48\textwidth}
        \centering
        \includegraphics[width=0.8\textwidth]{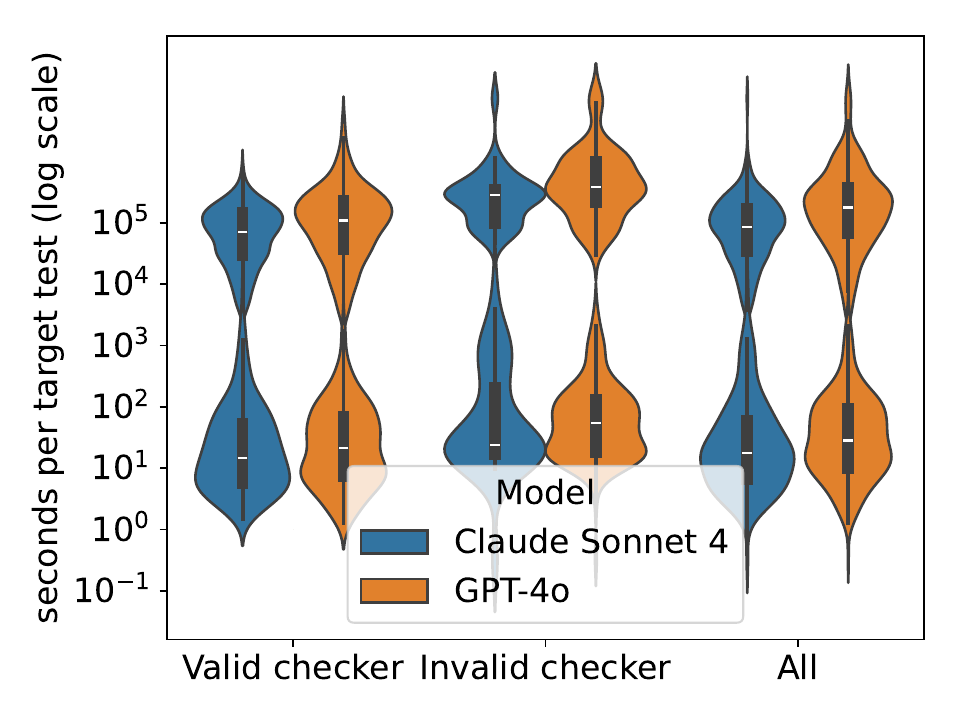}
        \caption{Time.}
        \label{fig:time}
    \end{subfigure}
    \hfill
    \begin{subfigure}[b]{0.48\textwidth}
        \centering
        \includegraphics[width=0.8\textwidth]{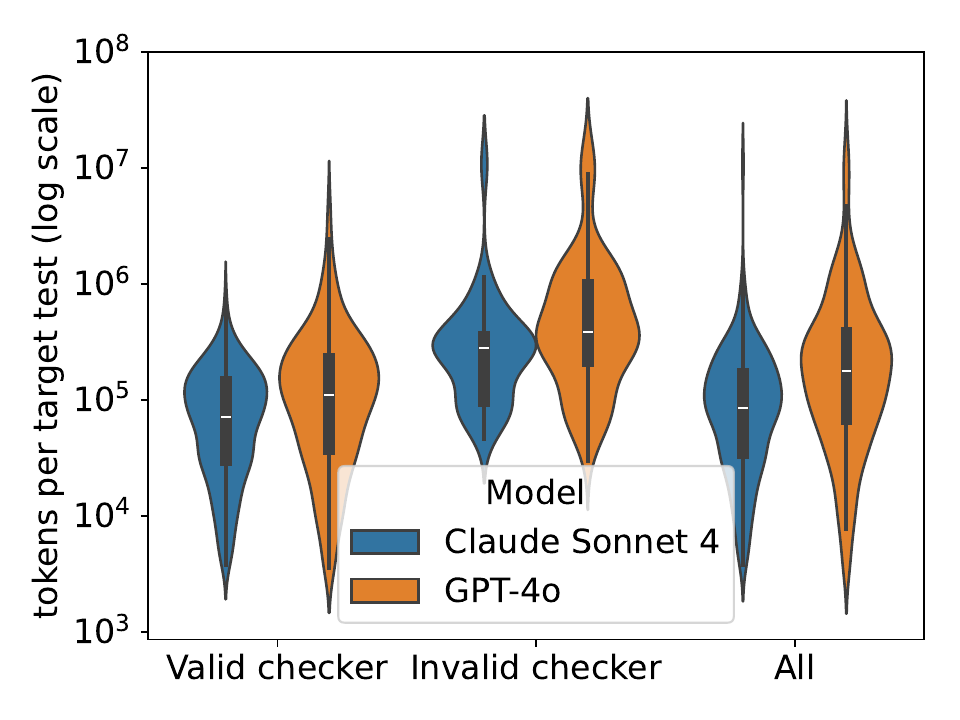}
        \caption{Tokens.}
        \label{fig:tokens}
    \end{subfigure}
    \caption{Distribution of costs to generate a runtime checker with \name{}.}
    \label{fig:costs}
\end{figure}

We measure the following costs imposed by \name{} to produce a runtime checker: 
(i) time to produce a runtime checker;
(ii) number of tokens consumed by queries to the LLM; and
(iii) monetary costs associated with the token consumption, based on OpenAI's and Anthropic's pricing as of October 2025.
In addition, to assess cost of using the checkers, we also measure
(iv) runtime overhead incurred when running the checkers.

Figure~\ref{fig:costs} presents the costs to generate a checker per target test case.
The median time to generate a checker that is successfully validated is \avgTimeToGenerateChecker{} with Claude Sonnet~4 and 21 seconds with GPT-4o.
However, for some outliers, the checker generation process can take several hours.
The upper ``belly'' in the distributions in Figure~\ref{fig:time} is from Cassandra, a particularly slow system in our evaluation.
Considering token costs, \name{} with Claude Sonnet~4 consumes 178k input tokens and produces 5k output tokens, on average per target test. This corresponds to a total of \avgTokensToGenerateChecker{} tokens and a monetary price of USD 0.60 per target test. With GPT-4o, \name{} consumes 552k input tokens and produces 9k output tokens.
This is a total of 561k tokens and a monetary price of USD 1.8 per test.
These results demonstrate that a more effective model tends to produce more valid checkers in less time, whereas a less effective model keeps trying to refine invalid checkers, consuming more resources.

To measure the runtime overhead imposed by using checkers produced by \name{}, we run the test files containing the target tests that \name{} was able to produce validated checkers for.
We run each test file five times with and without the checkers active in the system to account for variability in execution time and report average times.
We apply this process to three systems: Zookeeper, Cassandra and HBase.
For Zookeeper, the average execution time without and with the checkers is 1,139 and 1,170 seconds, respectively.
That is, the relative overhead is only 2.7\%.
For Cassandra, the average execution time without and with the checkers is 104 and 122 seconds, respectively.
That is a relative overhead of 17.3\%.
For HBase, the average execution time without and with the checkers is 4,308 and 6,041 seconds, respectively.
That is, we observe a relative overhead of 40.3\%.
It is important to note that these overhead results can be seen as an upper bound, as we run only the test files containing the target tests, which are a small subset of all tests in the system.
When running the entire test suite, or a production run of the target systems, the overhead is expected to be lower, as most tests do not have checkers enabled.

\section{Threats to Validity}

A first limitation concerns the generalizability of our evaluation.  
\name{} is assessed on four large, well-tested Java systems, which provides a diverse yet ultimately narrow sample of software.  
The approach may behave differently on other programming languages, smaller projects, or systems with fewer or qualitatively different tests.  
Moreover, the random sampling of tests and context tests introduces potential bias, as some types of assertions or code paths may be over- or under-represented.  
To mitigate such bias, we randomly sample a relatively large number of tests (4x100).

Another threat lies in the evaluation methodology.  
The bug-finding experiment relies on mutation testing as a proxy for real defects and uses only a single subject system, which limits the conclusions that can be drawn about real-world effectiveness.  
While mutants provide a controlled setting to measure fault detection, they may not fully reflect realistic fault distributions or semantics.  
Furthermore, the approach depends on large language models whose outputs vary with model version, temperature, and prompt phrasing, introducing nondeterminism.  
We mitigate this threat by documenting model versions and prompts, and by sharing detailed logs of our experiments.  

\section{Related Work}

\textit{Silent failures}.
Modern software systems experience increasingly complex failure 
patterns~\cite{Redundancy2017FAST,GrayFailureHotOS2017,Asynchrony2018OSDI,CrashTunerSOSP2019,
CorrelatedNSDI2020,DUPCheckerSOSP2021,qian2023vicious,Subtleties2006NSDI,Get2016ATC},  
such as partial failures~\cite{Watchdog2020NSDI}, fail-slow faults~\cite{GunawiFAST2018,yoo2021slow,SlowFaultStudy2025NSDI},
metastable failures~\cite{MetastableOSDI2022}.
Unlike failures that raise explicit error signals,
silent failures~\cite{wang2023silent,lou2025deriving} lack clear indicators and are therefore difficult to detect.
Lou et al.\ conduct a study~\cite{lou2022demystifying} on silent failures in distributed systems,
confirming that silent failures occur frequently and have significant consequences in practice.

\textit{Runtime checking}.
To detect silent failures, researchers have proposed 
frameworks~\cite{WiDS2007NSDI,Quinn2022omni,
PQL2005OOPSLA,EndoScope2008VLDB, guan2025faster, guan2025tracemop,
Arnold2008, Burnim2011a, Shen2025, Li2020b} that enable
developers to conveniently specify semantic properties and verify them 
at runtime. 
Early work enables developers to write parametric specifications and translates them into AspectJ-checkable formats~\cite{Chen2007} and enables to check type-state properties within a fixed overhead budget~\cite{Arnold2008}.
Recent work introduces a new specification languages~\cite{dawes2024checking} and extends runtime checking to new domains and languages~\cite{fse2022-DynaPyt,fse2025-Dylin}
A drawback of these approaches is that they require developers to manually write specifications, which is time-consuming and error-prone.
Instead, \name{} automatically generates runtime checkers from test cases, i.e., a resource available for most complex software.

\textit{Invariant mining}.
Manually writing specifications is time-consuming and error-prone.
A substantial body of work~\cite{DIDUCE2002ICSE,DySy2008ICSE,Detecting2009SOSP,CSight2014ICSE,
Burnim2010,Lu2007,mining2002popl,miningmalicious2007fse,miningparam2011icse} has explored automatically mining likely invariants from software execution traces.
Early systems, such as Daikon~\cite{Ernst1999ICSE} and Dinv~\cite{grant2018inferring}, 
infer key state relationships from test executions.
The more recent Oathkeeper~\cite{Oathkeeper2022OSDI} work extracts semantic relations between events by comparing test runs of buggy and patched versions.
These systems often suffer from high inaccuracy due to their statistical nature, and their 
inferred invariants are often not expressive enough to
capture the full richness of a system's semantics.
Our work differs in three key aspects.
First, we do not rely on execution traces, but instead leverage the source code of tests.
Second, instead of extracting invariants or patterns that fit specific templates, \name{}-generated checkers can contain arbitrary logic and state representations.
Finally, our approach reasons via LLMs about semantic properties encoded in tests, enabling it to generalize beyond the specific workloads and assertions encoded in the tests.  

\textit{Test transformation}.
Test cases serve as a valuable resource for discovering underlying program semantics and 
generating checkers.
Existing works such as PCheck~\cite{PCheck2016OSDI}, Ctests~\cite{Ctest2020OSDI} and 
ZebraConf~\cite{zebraconf2021eurosys} transform tests to executable checks to examine
validity and correctness of configurations.
In contrast, our work focuses on checking runtime behaviors instead of static configuration values.

\textit{LLM-based reasoning and code generation}.
The use of LLMs to reason about software and generate code has 
rapidly become a popular trend among researchers and 
developers~\cite{Barke2023,LLMRCACopilotEuroSys24,LLMRCAARCA2025EuroMLSys,LLMRCAReActFSE24,
LLMRCARCAgentCIKM24,LLMRCASOPFlowWWW25}.
TiCoder~\cite{TICODER2024tse} proposes an interactive workflow to partially formalize 
natural language guides and generates more accurate code.
ClassEval~\cite{ClassEval2024icse} proposes a new benchmark and evaluates
11 state-of-the-art LLMs on class-level code generation.
KNighter~\cite{knighter} automatically synthesizes static analyzers from 
historical bug patterns to expose new bugs.
Our system targets at runtime checker generation and deals with distinct challenges such 
as minimizing runtime safety risks.
\section{Conclusion}

This paper introduces \name{}, a novel approach that automatically infers semantic runtime checkers from existing tests by combining LLM-based synthesis, static analysis, and dynamic validation.
By leveraging large language models to generalize the intent behind test assertions and maintaining a shadow state for reasoning about system behavior, \name{} generates stateful and robust checkers that extend beyond the specific workloads encoded in tests.
Our evaluation on four complex software systems shows that \name{} produces substantially more correct and general checkers than prior work and detects significantly more otherwise missed errors.
These results demonstrate that LLM-guided synthesis, coupled with iterative validation, can transform existing tests into powerful runtime checkers, thereby enabling the broader and more practical use of runtime monitoring to detect silent failures in real-world software.

\section{Data Availability}

The code and data associated with our work is available:
\url{https://github.com/biabs1/FlyCatcher}.

\bibliographystyle{ACM-Reference-Format}
\bibliography{references,referencesMichael,referencesChang}

\end{document}